\documentclass[twocolumn]{aastex631}

\newcommand\xmm{\textit{XMM-Newton}}
\newcommand\chandra{\textit{Chandra}}

\newcommand\nicer{\textit{NICER}}

\newcommand\erosita{\textit{eROSITA}}

\newcommand\delcstat{$\Delta$C-stat}
\newcommand\pcm{cm$^{-2}$}

\newcommand\logxi{$\log (\xi$/erg~cm~s$^{-1})$}

\newcommand\source{ASASSN-20qc}
\newcommand\ftnli{ASASSN-14li}

\shorttitle{Multiphase outflow in ASASSN-20qc}
\shortauthors{Kosec et al.}
\graphicspath{{./}{figures/}}

\begin{document}

\title{Discovery of a variable multi-phase outflow in the X-ray-emitting tidal disruption event \source}

\correspondingauthor{P. Kosec}
\email{pkosec@mit.edu}

\author{P. Kosec}
\affiliation{MIT Kavli Institute for Astrophysics and Space Research, Massachusetts Institute of Technology, Cambridge, MA 02139}

\author{D. Pasham}
\affiliation{MIT Kavli Institute for Astrophysics and Space Research, Massachusetts Institute of Technology, Cambridge, MA 02139}

\author{E. Kara}
\affiliation{MIT Kavli Institute for Astrophysics and Space Research, Massachusetts Institute of Technology, Cambridge, MA 02139}

\author{F. Tombesi}
\affiliation{Physics Department, Tor Vergata University of Rome, Via della Ricerca Scientifica 1, 00133 Rome, Italy}
\affiliation{INAF – Astronomical Observatory of Rome, Via Frascati 33, 00040 Monte Porzio Catone, Italy}
\affiliation{INFN - Rome Tor Vergata, Via della Ricerca Scientifica 1, 00133 Rome, Italy}
\affiliation{Department of Astronomy, University of Maryland, College Park, MD 20742, USA}
\affiliation{NASA Goddard Space Flight Center, Code 662, Greenbelt, MD 20771, USA}



\begin{abstract}

Tidal disruption events (TDEs) are exotic transients that can lead to temporary super-Eddington accretion onto a supermassive black hole. Such accretion mode is naturally expected to result in powerful outflows of ionized matter. However, to date such an outflow has only been directly detected in the X-ray band in a single TDE, \ftnli. This outflow has a low velocity of just a few 100 km/s, although there is also evidence for a second, ultra-fast phase. Here we present the detection of a low-velocity outflow in a second TDE, \source. The high-resolution X-ray spectrum reveals an array of narrow absorption lines, each blueshifted by a few 100 km/s, which cannot be described by a single photo-ionization phase. For the first time, we confirm the multiphase nature of a TDE outflow, with at least two phases and two distinct velocity components. One highly ionized phase is outflowing at $910^{+90}_{-80}$ km/s, while a lower ionization component is blueshifted by $400_{-120}^{+100}$ km/s. We perform time-resolved analysis of the X-ray spectrum and detect that, surprisingly, the mildly ionized absorber strongly varies in ionization parameter over the course of a single 60 ks observation, indicating that its distance from the black hole may be as low as $400$ gravitational radii. We discuss these findings in the context of TDEs and compare this newly detected outflow with that of ASASSN-14li.

\end{abstract}



\section{Introduction} \label{sec:intro}

A tidal disruption event (TDE) is an exotic transient during which a star is disrupted as it ventures too close to a supermassive black hole \citep{Rees+88}. A significant fraction of the star's mass is accreted, which can lead to temporary super-Eddington accretion rates onto the black hole. In recent years, dozens of TDEs were discovered, in the optical band \citep{vanVelzen+21, Hammerstein+23} as well as in the X-rays \citep{Sazonov+21}. For recent reviews, see \citet{Gezari+21} and \citet{Saxton+20}. These events open a unique window into the lives of supermassive black holes in galaxies, the majority of which are inactive and thus challenging to study through other means. 

The violent, supercritical nature of this phenomenon is naturally expected to result in massive and high-velocity ($\sim0.1c$) outflows of ionized matter from the accretion flow \citep{Shakura+73}. Such outflows are observed in simulations of supercritical flows \citep{Ohsuga+09, Ohsuga+11, Takeuchi+13} and were observationally confirmed in other super-Eddington or highly accreting systems such as ultraluminous X-ray sources \citep{Pinto+16, Kosec+18b, Pinto+21} and active galactic nuclei \citep[e.g.][]{Pounds+03, Tombesi+10}. Physical models and simulations of super-Eddington accretion predict the existence of a geometrically and optically thick accretion flow \citep{Ohsuga+05}. The appearance and emission pattern of such a flow is thus inherently strongly non-isotropic. This is a possible explanation for why some TDEs are primarily bright in the optical band, and some others instead in the X-rays \citep{Dai+18}. It is thus of great importance to detect and understand the properties of outflows launched by TDEs, as they may strongly modify the inner accretion flow properties.

Observational evidence for these outflows in TDEs has however been sparse so far. The first TDE with a confirmed outflow in the X-rays was the nearby event \ftnli\ \citep{Miller+15}. The properties of this detected outflow phase are rather puzzling - \citet{Miller+15} found that the outflow had a  low systematic velocity of just $200-300$ km/s, in contrast with the expected velocities in excess of 5\% of the speed of the light, as seen in other supercritical systems mentioned above. \citet{Kara+18} later found evidence for a second, high-velocity component of the outflow at $\sim0.2$c in \ftnli, and \citet{Kara+16} found evidence for a high-velocity outflow originating from the inner accretion flow of the jetted TDE Swift J1644+57 through X-ray reverberation. However, it is unclear what is the driving mechanism of any of these X-ray outflows, and what is their relationship with the TDE. Clearly, the detection of outflows in further TDEs is necessary to understand their nature, physics, and impact on the accretion flow as well as the black hole surroundings. Outside the X-ray band, an outflow was detected in the TDE AT2019qiz using the optical line shape evolution \citep{Nicholl+20}, and likely corresponds to the expanding TDE photosphere, which is the dominant source of the optical and UV radiation.

Here we study \source\ \citep[z=0.056,][]{Stanek+20, Hinkle+22}, previously a low-luminosity AGN, which turned into an X-ray bright TDE (Pasham et al. submitted), sharing a number of similarities with \ftnli. \source\ was the target of a large multi-wavelength campaign in 2021. It was detected in the X-rays by the \erosita\ survey and observed by \xmm\ and \nicer\ observatories. Its X-ray spectrum reveals a soft X-ray continuum, which can be broadly described by a disk blackbody with a temperature of $\sim$90 eV. Similar to ASASSN-14li \citep{Kara+18}, the spectrum appears to show a significant broad dip around the Wien tail. This feature, if modeled as an absorption line, suggests an outflow velocity of 0.3c. We refer to Pasham et al. (submitted) for a detailed analysis of this ultra-fast outflow component.

In this paper, we instead focus on the high-resolution \xmm\ RGS spectra of \source, which reveal an array of absorption lines, similar to those found in ASASSN-14li. \source\ is only the second TDE to exhibit such narrow lines. The structure of this paper is as follows. Our data preparation and reduction is summarized in Section \ref{sec:data}. The spectral modelling and the results are described in Section \ref{sec:results}. We discuss the results and their implications in Section \ref{sec:discussion} and conclude in Section \ref{sec:conclusions}.

\section{Data Reduction and preparation} \label{sec:data}

\xmm\ \citep{Jansen+01} observed \source\ six times to date. Here we analyze observation 0852600301, the only observation which occurred when the source was in a high flux state, allowing us to use the high-spectral resolution data from the Reflection Grating Spectrometers (RGS). During the remaining five \xmm\ observations, \source\ reached an order of magnitude lower count rates. Observation 0852600301 occurred on March 14 2021 and had a duration of $\approx$60 ks. The data were downloaded from the XSA archive and reduced using a standard pipeline with \textsc{sas} v20, \textsc{caldb} as of 2022 April. We use data from RGS \citep{denHerder+01} and from European Photon Imaging Camera (EPIC) pn \citep{Struder+01}.

The RGS data were reduced following standard routines using the \textsc{rgsproc} procedure, centering the extraction regions on the coordinates of \source. We filtered for any periods of high background, but the RGS detectors were not significantly affected by any major flares, and the clean exposure of each detector is about 58.5 ks. RGS 1 and 2 data were not stacked, but were fitted simultaneously in all spectral fits using a cross-calibration constant. The value of this parameter was always close to 1, indicating $<5$\% calibration difference between the two instruments. We binned the RGS spectra by a factor of 3 to achieve only mild oversampling of the instrumental spectral resolution. This was achieved with the `bin' command in the \textsc{SPEX} fitting package \citep{Kaastra+96}. We use the RGS data in the 15 \AA\ (0.83 keV) to 36 \AA\ (0.34 keV) wavelength range. The lower limit is set by the data quality - there is no signal in RGS below 15 \AA\ due to the softness of the \source\ X-ray spectrum.

To constrain the continuum model, we also examined the EPIC PN data that extends to higher energies. The EPIC PN instrument was operated in the Small Window mode. The data were reduced using the \textsc{epproc} procedure, and only events of PATTERN$\leq$4 (single/double) were accepted. We screened for background flares with a threshold of 0.15 ct/s in the 10-12 keV lightcurve, keeping in mind the small area of the active CCD during Small Window mode operation. This resulted in a clean exposure time of 35.5 ks. The source region was a circle with a radius of 15 arcsec centered on \source\ position. We specifically chose a small source region size to maximize the signal-to-noise ratio and decrease the background importance at higher energies ($>1$ keV) considering the extreme softness of \source. The background region was a polygon on the other side of the pn CCD as the source, at least 150 arcsec away. The source has a count rate of 3 ct/s, and so pile-up should not be an issue during the observation. We confirmed this by assessing the pile-up plots produced by the \textsc{epatplot} routine. The data were grouped using the \textsc{specgroup} procedure to at least 25 counts per bin and at the same time to oversample the instrumental resolution by at most a factor of 3. We use EPIC pn in the wavelength range between 8 \AA\ (1.55 keV, limited by data quality) and 15 \AA\ (0.83 keV, RGS data available above this limit). We do not use any EPIC pn data beyond 1.55 keV as the spectrum is strongly background-dominated in that range.

We fit the spectra in the \textsc{spex} \citep{Kaastra+96} fitting package. All reduced spectra were converted from \textsc{ogip} format into \textsc{spex} format using the \textsc{trafo} routine. We use Cash statistic \citep{Cash+79} to analyze the spectra. All uncertainties are provided at 1$\sigma$ significance.

\section{Spectral modelling and results} \label{sec:results}

\subsection{RGS analysis} \label{subsec:rgsfitting}

\begin{figure*}
\begin{center}
\includegraphics[width=\textwidth]{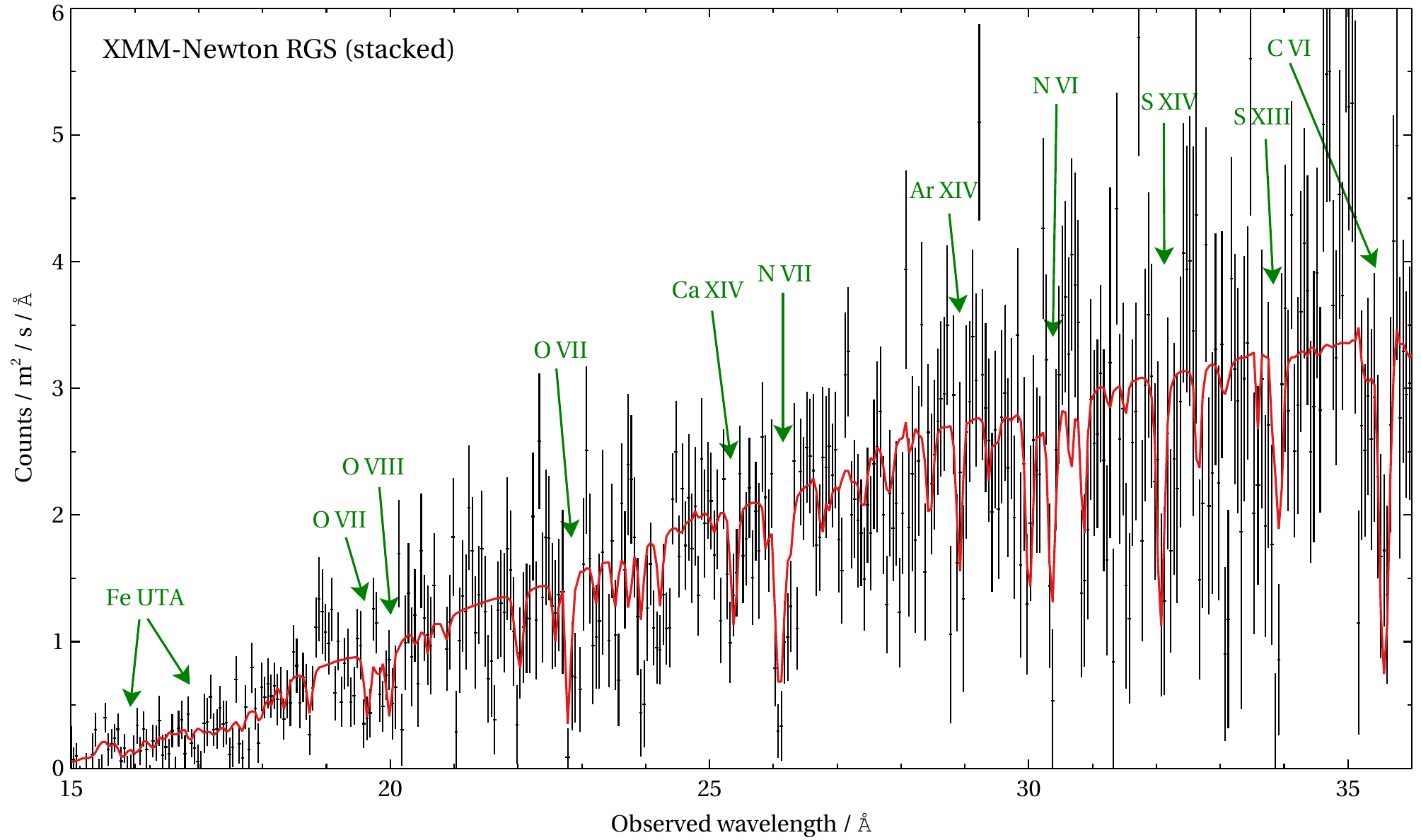}
\caption{\xmm\ RGS spectrum of \source. RGS 1 and RGS 2 spectra are stacked and over-binned for visual purposes only. The best-fitting ionized absorption model, consisting of two photo-ionized components is shown in red. The most notable elemental transitions are shown with green labels. Fe UTA denotes the unresolved transition array of Fe absorption lines. \label{RGS_spectrum}}
\end{center}
\end{figure*}

The RGS spectrum of \source, shown in Fig. \ref{RGS_spectrum} reveals a plethora of absorption lines which cannot be resolved with EPIC pn due to its poor spectral resolution below 1 keV. Many of the lines appear highly significant - we particularly note the absorption feature at 26 \AA. Most of the lines are narrow, showing widths of the order of a few 100s km/s (full width at half maximum) at most, indicating an ionized absorber with low velocity width.

We begin the spectral modelling with a baseline continuum fit. The broadband continuum is described with a disk blackbody, the \textsc{dbb} component within \textsc{spex}, with a temperature of around 0.18 keV. The definition of the disk blackbody temperature is different in \textsc{spex} and \textsc{xspec}, resulting in \textsc{spex} \textsc{dbb} temperatures being higher by roughly a factor of 2. The disk blackbody is redshifted by z=0.056 using the \textsc{reds} model. Finally, Galactic absorption is applied using the \textsc{hot} component. We fix the neutral column density to $1.2\times10^{20}$ \pcm\ \citep{HI4PI+16}. The final fit statistic of the continuum model is C-stat=1900.68 with 1190 degrees of freedom (DoF).

To determine the absorber properties, we add a \textsc{pion} photo-ionized absorption component. \textsc{pion} \citep{Miller+15, Mehdipour+16} self-consistently calculates absorption line strengths using the ionizing balance determined from the currently loaded spectral continuum model. The ionizing balance is calculated on the fly as the continuum changes during spectral fitting. The \textsc{pion} component allows us to recover the ionized absorber properties such as its column density, ionization parameter \logxi, outflow velocity and velocity width. The plasma elemental abundances are fixed to Solar values.

The addition of one \textsc{pion} component is highly significant ($>>5\sigma$ using F-test) and improves the fit quality to C-stat=1748.85 (\delcstat=151.83 over the baseline continuum for 4 additional DoF). It requires an outflow with a mild velocity of $\sim800$ km/s, and an ionization parameter \logxi\ of 3.5. However, many of the absorption lines in the RGS data are still not well fitted in this spectral fit with a single absorber. 

This could indicate that a single photoionized absorber is insufficient to fit the observed absorption lines. In other words, the outflow is likely multi-phase. To test this hypothesis, we add a second \textsc{pion} component to the previous spectral fit. This step again improves the spectral fit, now to C-stat=1669.58, a further \delcstat=79.27 over the single phase absorber fit (for 4 extra DoF), and \delcstat=231.1 over the baseline continuum fit (for 8 extra DoF in total). The best-fitting absorber and continuum parameters are listed in Table \ref{RGS_table}, and the spectral fit is shown in Fig. \ref{RGS_spectrum}. We find one highly ionized absorber with an ionization parameter \logxi\ of $3.75_{-0.22}^{+0.15}$, and a high column density of $0.25_{-0.09}^{+0.14} \times 10^{24} $ \pcm. The second component is much more mildly ionized at \logxi$\sim1.39_{-0.26}^{+0.25}$, and has a much lower column density of $1.5_{-0.3}^{+0.4} \times 10^{21} $ \pcm. Both show low velocity widths, and outflow velocities comparable to those of warm absorbers in regular AGN, but also comparable to the ionized absorber found in the TDE \ftnli\ \citep[with an outflow velocity of 100-500 km/s,][]{Miller+15}. The highly ionized component is significantly faster at $910_{-90}^{+80}$ km/s, while the second one has a velocity of $380_{-100}^{+120} $ km/s.

We note that the column density of the highly ionized absorber is very high (above $10^{23}$ \pcm), and has significant uncertainties. This value is more than $10\times$ higher than the column density found by \citet{Miller+15} in \ftnli. It is possible that this column density is incorrectly determined from the limited RGS spectrum. Specifically, the spectrum is lacking any information below $15$ \AA. This spectral region would be important in placing strong upper limits on the outflow column density and the ionization parameter, thanks to the many Fe transitions in the 8-12 \AA\ band. Unfortunately, \source\ is spectrally very soft and RGS does not offer sufficient collecting area at higher energies. In the following section, we will use simultaneous EPIC pn coverage at higher energies (0.8-1.5 keV) to obtain a more reliable measurement of the highly ionized absorber properties.

We further test the properties of the absorbers by freeing their covering fractions (fcov parameter in \textsc{spex}). No significant evidence is found for the covering fraction being lower than 1 for any of the two absorbing phases.

\begin{deluxetable}{ccc}
\tablecaption{Best-fitting properties of the continuum and the two ionized absorbers in the spectrum of \source, determined from an RGS-only analysis. \label{RGS_table}}
\tablewidth{0pt}
\tablehead{
\colhead{Component} & \colhead{Parameter} & \colhead{Value}  
}
\startdata 
disk & norm &  $(320 \pm 40)\times10^{16}$  m$^{2} $\\
blackbody  & kT & $0.197_{-0.003}^{+0.004} $ keV \\
\hline
highly ionized & N$_{H}$ & $2.5_{-0.9}^{+1.4} \times 10^{23} $ \pcm \\
absorber  & \logxi\ & $ 3.75_{-0.22}^{+0.15}$ \\
 & outflow velocity & $ 910_{-80}^{+90} $ km/s \\
 & velocity width & $  80 \pm 20$ km/s\\
 & \delcstat\  & $151.83 $ \\
\hline
mildly ionized&  N$_{H}$ & $1.5_{-0.3}^{+0.4} \times 10^{21} $ \pcm \\
absorber  & \logxi\ & $1.39_{-0.26}^{+0.25} $ \\
 & outflow velocity & $380_{-120}^{+100} $ km/s \\
 & velocity width & $240_{-60}^{+70} $ km/s\\
 & \delcstat\  & $79.27 $ \\
\enddata
\end{deluxetable}

We also tried adding a third photoionization phase to the spectral fit. This improves the statistic moderately to C-stat=1645.75, a \delcstat=23.83 fit improvement (for 4 extra degrees of freedom) over the previous spectral fit. The best-fitting absorber has a column density of $2.6_{-0.6}^{+0.8} \times 10^{20}$ \pcm, \logxi\ of $-1.15_{-0.17}^{+0.23}$, outflow velocity of   $630^{+130}_{-140}$ km/s and a velocity width of $200_{-80}^{+170}$ km/s. It improves the spectral fit particularly around 24 \AA\ (Fig. \ref{RGS_spectrum}). The evidence for the third phase indicates that the ionized outflow in \source\ is likely highly complex and strongly multi-phase. However, only the first two phases strongly modify the high-resolution spectrum and are unambiguously detected at high significance. For this reason, in all our following spectral fits we only include two low-velocity absorber phases.

Finally, we test for the presence of a broad ultra-fast outflow, found in the EPIC spectrum (Pasham et al. submitted), in the high-resolution RGS spectra. We take the baseline double slow absorber spectral fit and add a third \textsc{pion} component, now strongly broadened with an FWHM velocity width of about 70500 km/s (30000 km/s at 1$\sigma$, same as used in Pasham et al.), outflowing with a large velocity ($<0.1$c). The addition of such a component is not highly significant in the RGS data alone, with a fit statistic C-stat=1651.90, a \delcstat=17.68 fit improvement. This outcome is not surprising since the residual attributed to the UFO in EPIC pn data extends between 0.7 and 1.0 keV. RGS statistics are low in the 0.7-0.8 keV range and data are completely missing above 0.8 keV.

\subsection{Simultaneous RGS and EPIC pn spectral modelling}

The spectral analysis of the RGS data alone reveals many narrow absorption lines indicating the presence of a low velocity, multi-phase outflow in \source. At the same time, the RGS results are potentially limited - we are unable to reliably measure the outflow properties due to the lack of RGS data above 0.8 keV. To perform the best possible measurement, we need to combine the RGS and EPIC pn datasets and fit them simultaneously. 

Unfortunately, it is impossible to simply fit these spectra simultaneously over the full energy band. The pn data dominate the fitting statistic, with a count rate $>10\times$ higher than that of both RGS detectors summed. At the same, the pn data offer a much poorer spectral resolution, and all the individual narrow absorption lines are blended together. Therefore, the fit is driven by broad continuum-like shapes instead of the individual line positions, shapes and optical depths. This can lead to incorrect results, especially considering that the outflow is multi-phase. Furthermore, there are residual calibration differences between the RGS and EPIC pn instruments, which vary across the overlapping energy band of the instruments \citep{Detmers+10}. These can be important in bright sources such as \source\ (with small Poisson errorbars on individual EPIC pn data points) and can systematically skew the best-fitting outflow properties.
 
A common way to avoid this issue is to ignore the EPIC pn data in the energy band where RGS has sufficient statistics (around 0.8-0.85 keV in this study), and use a cross-calibration constant to account for any residual calibration uncertainties between the two instruments at the contact energy where the two datasets meet, fitting for the value of this constant. However, this cannot be done in the case of \source. The EPIC pn spectrum shows an absorption residual extending from 0.7 to 1.0 keV, i.e. the residual extends across the band where RGS loses statistics, and continues into the EPIC pn-only band. Therefore, it is possible that the cross-calibration constant in fact might fit the shape of the absorption residual at the point of contact between the two datasets, instead of broad instrumental normalization differences. For this reason it is challenging to use the two datasets simultaneously. On the one hand, we do not want to ignore RGS data below the absorption residual ($<0.7$ keV), thus losing all the spectral resolution in the important 0.7-0.8 keV region, but on the other hand the unknown cross-calibration difference between the two instruments at the contact point can introduce systematic error in the spectral fit.

Thankfully, the cross-calibration differences between RGS and EPIC pn across the energy band are unlikely to be too large, and are at most 15 \% \citep{Detmers+10}. We can therefore explore the parameter space of these possible differences, to see how much our uncertainty in understanding the instrument cross-calibration affects our outflow modelling and its best-fitting properties. We repeat the same spectral fit for different values of the cross-calibration constant, kept frozen in the range between 0.85 and 1.15, with a step size of 0.05 (7 spectral fits in total for each spectral model).

As mentioned in Section \ref{sec:intro}, a single blackbody is insufficient to describe the broadband X-ray (0.3-1.5 keV) spectrum of \source. To improve the description of the continuum, we use two different spectral models describing the combined dataset, which have previously been used to fit TDE X-ray spectra. 

The first model includes a disk blackbody, two slow absorbers producing the narrow absorption lines, and a fast UFO phase with a large velocity broadening (30000 km/s at 1$\sigma$ = 70500 km/s full width half maximum). This model (blackbody + UFO absorption) model was previously used by \citet{Kara+18} and is also employed to describe the EPIC-pn spectrum of \source\ by Pasham et al. (submitted). The model represents standard TDE disk blackbody emission from a compact accretion disk, modified by absorption from a high-velocity outflow, launched by the extreme mass accretion rate during the tidal disruption.

We also use an alternative model which does not contain any UFO absorption, and instead applies a continuum consisting of two regular blackbodies, obscured by the two low-velocity absorbers. A disk blackbody and additional warmer continuum has been employed in several recently discovered nuclear transients, including \ftnli\ \citep{Kara+18}, ASASSN-18el \citep{Ricci+20, Masterson+22} and in two TDE candidates later found to also exhibit quasi-periodic eruptions \citep{Miniutti+19, Chakraborty+21}. This model represents a physical scenario where two different physical components are responsible for the observed spectral continuum. One of these components could be a regular blackbody from an accretion disk, the second one could originate from shocks, or from Comptonization within a corona. This model is motivated by the broad spectral shape of the UFO model (with a large velocity broadening producing no discrete narrow features), which could in principle be reproduced by a more complex emission continuum spectral shape.

By applying these two different spectral models we are as agnostic as possible to the interpretation of the underlying spectral continuum (which is discussed elsewhere), and show that the determined low-velocity absorber properties are not dependent on this interpretation.

\begin{figure}
\begin{center}
\includegraphics[width=\columnwidth]{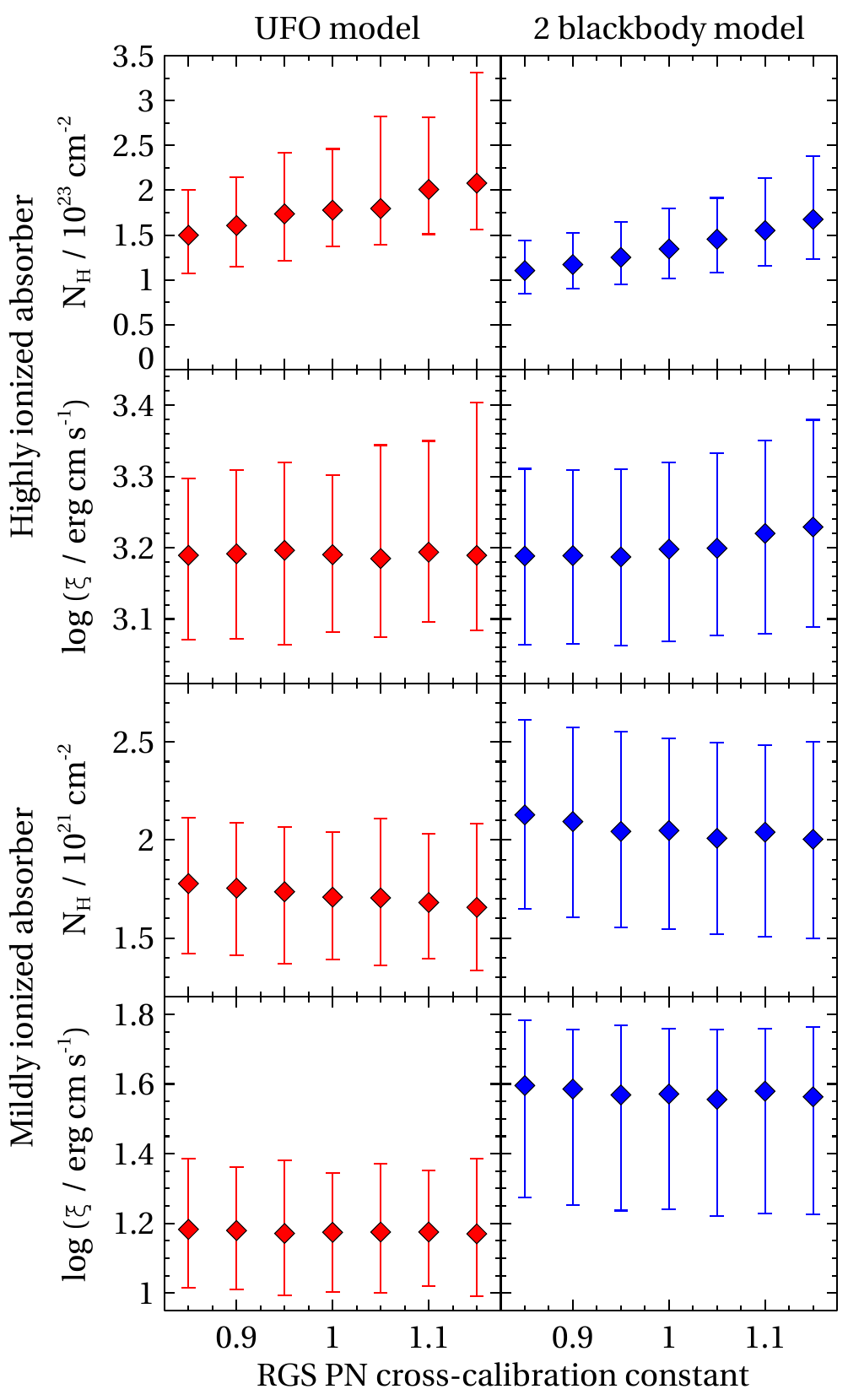}
\caption{The best-fitting properties of the low-velocity absorbers (assuming photo-ionization balance) versus the value of the cross-calibration constant between RGS and PN instruments. The left panels show the results for the UFO model, while the right panels show the double blackbody model. The top two panels of each column show the column density and ionization parameter for the highly ionized absorber, while the lower two panels show the column density and the ionization parameter for the mildly ionized absorber phase. \label{RGS_EPIC_fitting}}
\end{center}
\end{figure}

\begin{figure}
\begin{center}
\includegraphics[width=\columnwidth]{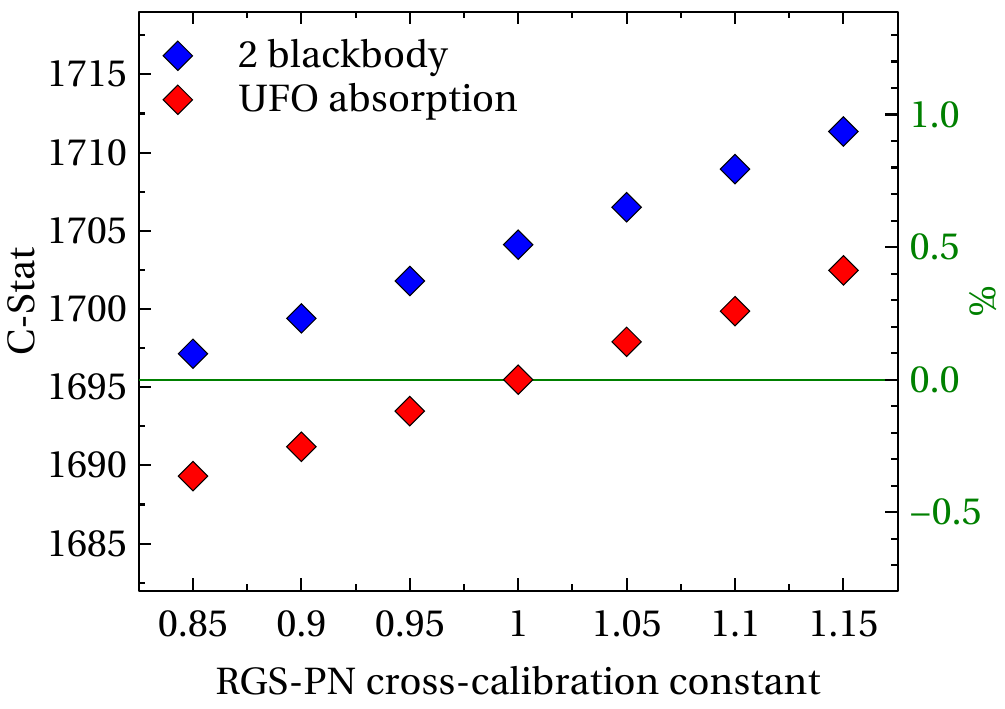}
\caption{The C-stat fit statistic for each model versus the assumed value of the RGS-PN cross-calibration constant. The UFO models are shown in red, while the double blackbody models are in blue. The right Y-axis shows the relative C-stat difference (in \%) from the best-fitting UFO model assuming perfect RGS-PN cross-calibration. \label{RGS_EPIC_fitting_Cstat}}
\end{center}
\end{figure}

The absorber properties versus the value of the cross-calibration constant are shown in Fig. \ref{RGS_EPIC_fitting}. We find no large steps in the absorber properties over the explored range of cross-calibration constants. All parameters are varying smoothly, with no sudden jumps. Similarly, the fit quality C-stat does not vary significantly with the cross-calibration constant, spanning at most \delcstat$\sim15$ across the studied range of cross-calibration constant values (Fig. \ref{RGS_EPIC_fitting_Cstat}). In the UFO spectral model, the best-fitting UFO component outflow velocity is around 0.32c, fully consistent with the results of Pasham et al. (submitted). It does not vary significantly with the RGS-PN cross-calibration constant. These investigations give us confidence that our results on the properties of the multi-phase warm absorber are robust to uncertainties in the continuum model and the cross calibration constant.

We find that the best-fitting parameters of the highly ionized absorber are consistent between the two spectral models, and do not vary very strongly with the RGS-PN cross-calibration constant. Conservatively, we can therefore conclude that the best-fitting highly ionized absorber column density is in the range of $(0.9-3.3)\times 10^{23}$ \pcm, and its ionization parameter \logxi\ in the range of $3.0-3.4$. We note that these limits are not to be taken at 1$\sigma$ confidence given the unknown exact value of cross-calibration between RGS and pn. These results, especially the ionization parameter, are somewhat lower than the absorber properties recovered from RGS alone, confirming that RGS over-estimates the highly ionized absorber, most likely due to its lack of signal in the important wavelength band below 15 \AA. The outflow velocity and the velocity width is consistent with the RGS-only analysis, and does not vary with the RGS-PN cross-calibration constant.

We find larger differences between the two models when comparing the mildly ionized absorber. The UFO model results in a somewhat lower column density as well as the ionization parameter. We conclude that the column density of the absorber is most likely in the range of $(1.3-2.6)\times 10^{21}$ \pcm, and its ionization parameter \logxi\ is in the range of $1.0-1.8$. We find that the best-fitting parameters do not vary with the RGS-PN cross-calibration constant. The best-fitting outflow velocity and velocity width are again fully consistent with the RGS-only spectral analysis, and do not vary with the cross-calibration constant.

The total C-stat for both spectral models is between 1689 and 1711. The UFO model has 1190 degrees of freedom (DoF), while the double blackbody model has 1192 DoF. We find that the UFO model is preferred to the double blackbody model by about \delcstat$\sim8-9$ depending on the exact value of the cross-calibration constant. Simply comparing the difference in DoF and in C-stat values, the UFO model is preferred by our \xmm\ data by $\sim2.5\sigma$. However, this simplified comparison does not take into account the great number of DoFs in both of these models and is thus only a very rough estimate.

\subsection{Time-resolved spectral analysis}

\citet{Miller+15} found that the ionized absorption in \ftnli\ is variable on the timescale of a single $\sim100$ ks \xmm\ observation. The detection of such fast variability puts strong upper limits on the distance of the ionized outflow from the black hole. Below we investigate whether variability can be detected on single observation timescales in \source\ as well. As a first order approach, we split the \xmm\ observation into two segments of equal exposure and perform the spectral fit again, tying certain parameters (which are unlikely to vary) together. Given the difficulties discussed above in the spectral model choice and the cross-normalization value when analyzing RGS and EPIC pn data together, we perform the outflow variability test for the RGS data alone. Because we are only fitting the RGS data, the parameters recovered for the highly ionized outflow phase might be less trustworthy than in the analysis above, however the variability (in the narrow absorption lines) would be detected nevertheless.

We fit both observation segments simultaneously, employing the two phase outflow spectral model from Section \ref{subsec:rgsfitting}. We allow the disk blackbody properties to vary between the segments, as well as the absorber column densities and ionization parameters. We couple the outflow velocities and velocity widths as they are unlikely to significantly change between the two segments, and since we do not visually observe any apparent line shifts between the segments.

\begin{figure}
\begin{center}
\includegraphics[width=\columnwidth]{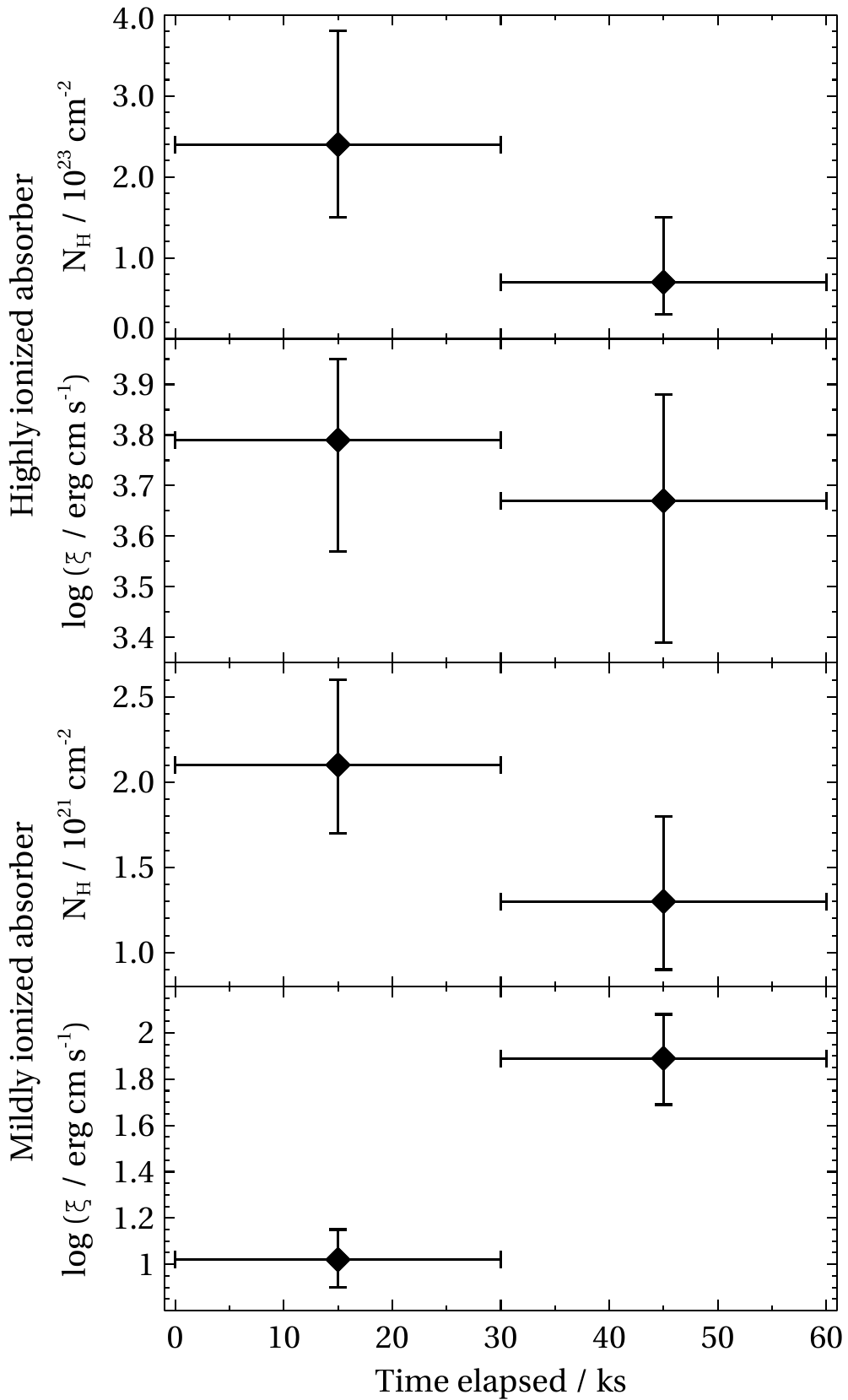}
\caption{Variation of the best-fitting properties of the low-velocity absorbers during the 60 ks \xmm\ observation, investigated by splitting it into two segments. The top two panels show the column density and the ionization parameter for the highly ionized absorber phase, while the lower two panels show the column density and the ionization parameter for the mildly ionized phase. \label{RGS_time_resolved_plot}}
\end{center}
\end{figure}

The fitting results are shown in Fig. \ref{RGS_time_resolved_plot} and in Table \ref{RGS_time_resolved_table}. We find that some of the absorber parameters change significantly between the two segments. The largest variation ($>3\sigma$) is surprisingly seen in the mildly ionized absorber ionization parameter \logxi, which changes from $1.02_{-0.12}^{+0.13} $ to $1.89_{-0.20}^{+0.19}$ over the course of the 60 ks \xmm\ observation. We also detect possible variability in the mildly ionized absorber column density ($\sim1.5\sigma$ significance), and in the highly ionized absorber column density ($\sim2\sigma$ significance). Finally, a large difference is also observed in the disk blackbody normalization, however this is necessarily correlated with the apparent variation in the highly ionized absorber column density (change in continuum absorption is significant as the $N_{H}$ value is more than $10^{23}$ \pcm\ in one of the segments). The observation with the higher highly ionized absorber column density (and greater blackbody luminosity) also shows a slightly greater ionization parameter \logxi, however its variation is not statistically significant.

\begin{deluxetable*}{cccc}
\tablecaption{Time-resolved analysis of the ionized absorption in \source. Best-fitting properties of the continuum and the two ionized absorbers, recovered by fitting RGS data only. The full \xmm\ observation was split into two segments with roughly equal exposure. \label{RGS_time_resolved_table}}
\tablewidth{0pt}
\tablehead{
\colhead{Component} & \colhead{Parameter} & \colhead{Segment 1} & \colhead{Segment 2}  
}
\startdata 
disk& normalization & $(390 \pm 50)\times10^{16}$  m$^{2} $& $(280 \pm 30)\times10^{16}$  m$^{2}$\\
blackbody& kT & $0.196_{-0.004}^{+0.005} $ keV& $0.194 \pm 0.004$ keV\\
highly ionized& N$_{H}$ & $2.4_{-0.9}^{+1.4}\times10^{23} $\pcm& $ 0.7_{-0.4}^{+0.8}\times10^{23} $ \pcm\\
absorber&\logxi & $3.79_{-0.22}^{+0.16} $& $3.67_{-0.28}^{+0.21}$\\
& outflow velocity & \multicolumn{2}{c}{$910 \pm 90$ km/s} \\ 
& velocity width & \multicolumn{2}{c}{$90 \pm 20$ km/s}\\
mildly ionized& N$_{H}$ &  $2.1_{-0.4}^{+0.5}\times10^{21} $\pcm& $1.3_{-0.4}^{+0.5}\times10^{21} $\pcm\\ 
absorber&\logxi &$1.02_{-0.12}^{+0.13} $& $1.89_{-0.20}^{+0.19}$\\ 
& outflow velocity &\multicolumn{2}{c}{$410 \pm 90 $ km/s}\\ 
& velocity width &\multicolumn{2}{c}{$270 \pm 50 $ km/s}\\ 
\enddata
\end{deluxetable*}

Finally, we consider the variation of the absorber outflow velocities and velocity widths between the two observation segments. We untie all the spectral parameters of the two phase model and re-fit. We found that both the outflow velocities and velocity widths of both absorbers are consistent at $1\sigma$ confidence between the two segments.

\section{Discussion} \label{sec:discussion}

We study the RGS and PN spectra of the TDE \source\ and significantly detect a multi-phase, low-velocity ionized absorber. Two distinct velocity and ionization components are confirmed, but with further evidence for a third ionization phase, the true nature of this outflow is likely much more complex. The highly ionized phase is faster at 900 km/s, while the mildly ionized phase is outflowing with a velocity of 400 km/s. Both of these values are similar to the outflow detected in \ftnli\ with a velocity of $100-500$ km/s \citep{Miller+15}.

The ionization parameter of the highly ionized phase, \logxi\ of $3.0-3.4$ is comparable with the ionized outflow of \ftnli, but it has a much higher column density of $\sim10^{23}$ \pcm\, versus $(0.5-1.3)\times10^{22}$ \pcm\ in \ftnli. Such a high column density is surprising, and more similar to the column densities measured in ionized obscurers \citep[e.g.][]{Partington+23} and UFOs \citep[e.g.][]{Tombesi+11} in AGN. However, those outflows show much higher systematic velocities than observed in \source, with obscurers reaching thousands km/s, and UFOs exceeding 0.05c. Nevertheless, the absorber of \source\ is relatively highly ionized, and so it is still transparent to even soft X-rays, as opposed to obscurers seen in AGN, which tend to absorb most of the source soft X-ray continua.

The mildly ionized absorber phase (alongside with the possible third absorption phase) with an ionization parameter of around 1.5 and a column density of around $10^{21}$ \pcm\ is novel and has not been previously detected in TDEs. In its properties, it is very similar to warm absorbers in AGN \citep{Blustin+05, Laha+14}. The best-fitting time-averaged properties of this phase might suggest that it is a remnant outflow from previous black hole activity, only re-illuminated by the current TDE outburst. However, the fast time variability of this component argues against such interpretation.

We can use the best-fitting ionization properties (assuming photo-ionization equilibrium) of the two phases to estimate their distance from the black hole. Following \citet{Kosec+20a}, we can use the definition of the ionization parameter \logxi\ and the definition of the outflow column density $N_{\rm H}$ to get the following expression for the distance $R$ of the outflow from the ionizing source:

\begin{equation}
    R = \frac{L_{\rm{ion}}}{N_{\rm{H}}\xi} \frac{\Delta R}{R}
\end{equation}

where $L_{\rm ion}$ is the 1-1000 Ryd ionizing luminosity, and $\Delta R$ is the thickness of the absorbing layer. By taking $\frac{\Delta R}{R}=1$ as the relative thickness of the absorbing layer cannot be larger than unity, we can estimate the maximum distance of the absorber from the black hole. The ionization luminosity of \source, from our X-ray spectral fits, is about $5\times 10^{44}$ erg/s. The maximum distance for the highly ionized phase is thus $2\times10^{18}$ cm = 0.6 pc and for the mildly ionized phase is around $1.1\times10^{22}$ cm = 4000 pc. We convert these into gravitational radii (R$_{\rm G}$) units assuming a black hole mass of $3\times10^7$ $M_{\odot}$ \citep[Pasham et al. submitted, we note that][estimated a similar black hole mass of $2\times10^7$ $M_{\odot}$]{Hinkle+22}. For the highly ionized phase, we calculate a maximum distance of $4\times10^5$ R$_{\rm G}$ from the black hole, and for the mildly ionized phase we estimate $2\times10^9$ R$_{\rm G}$. These are pc-scale distances, and serve as an absolute upper limit on the location of the absorbers given their ionization properties. If the absorbers have low relative thicknesses ($\frac{\Delta R}{R}<<1$, i.e. low volume filling factors), they will be located much closer to the black hole than our estimates.

Instead, we could assume that the outflow velocity of each phase is comparable with the escape velocity at its location. By making this assumption, the distance of the outflow from the black hole is:

\begin{equation}
    R = \frac{2GM}{v^2} = 2\frac{c^2}{v^2} R_{\rm G}
\end{equation}

where $R_{\rm G}$ is the gravitational radius of the black hole. This assumption results in a distance of $\sim2\times10^5$ $R_{\rm G}$ (0.4 pc) for the highly ionized phase, and a distance of $\sim10^6 R_{\rm G}$ (2 pc) for the mildly ionized absorber phase. The velocity distance estimate is similar to the upper limit from the ionization parameter for the highly ionized component, but wildly different for the mildly ionized component. This result suggests that the mildly ionized component indeed has a very low relative thickness $\frac{\Delta R}{R}<10^{-3}$, and thus a low volume filling factor.

Importantly, we also detect significant time variability over the course of the single \xmm\ observation. Similar variation was previously detected also in \ftnli. Surprisingly, the statistically strongest variation is detected in the mildly ionized absorber properties, rather than in the highly ionized absorber. However, this may be due to the current dataset quality, where the variation in the mildly ionized absorber lines is more easily detected than the variation in the highly ionized absorber.

This variation puts an important upper limit on where the outflow can physically reside. If the outflow transverse velocity (with respect to the X-ray source) is too low, it is unable to cross the X-ray source in the limited time (here $\sim$30 ks) between the two segments of our observation. Recent results on the size of X-ray emitting regions show that the region is very compact, in most cases with a radius smaller than 10 $R_{\rm G}$ \citep{Morgan+08, Dai+10, Sanfrutos+13, Chartas+16}. We note that if the emitting region is in fact larger, the resulting distance of the absorber from the black hole is even smaller than the estimate below. Conversely, if the X-ray emitting region is smaller, the absorber can be located farther from the black hole. 

For the \source\ black hole mass estimate ($3\times10^7$ $M_{\odot}$), the radius of 10$R_{\rm G}$ is $4 \times 10^{13}$ cm. To cross this radius in 30 ks (or 60 ks for the full diameter) and introduce time variability in the ionized absorption, the absorber needs to move with a transverse velocity of at least $1.5 \times 10^{9}$ cm/s = 15000 km/s. To estimate a rough distance of this absorber from the black hole, we assume that the transverse velocity of the absorber is comparable with the Keplerian velocity at its location. Then its distance from the black hole is roughly:

\begin{equation}
    R = \frac{GM}{v^2} = \frac{c^2}{v^2} R_{\rm G} = 400 R_{\rm G}
\end{equation}

The time variability puts a much stronger upper bound on the location of the absorber than the ionization balance and the outflow velocity estimates. It indicates that the absorber cannot be a remnant warm absorber from the previous black hole activity, located at pc scales away from the X-ray source. The outflow hence most likely originates by some launching mechanism from the TDE. \citet{Miller+15} reached a similar conclusion based on the properties of the outflow in \ftnli.

We note that given the black hole mass and the X-ray emitting region size estimates, the outflow has unusual velocity components - a very high toroidal component (15000 km/s) and a very low line-of-sight velocity (400-900 km/s). If the assumptions of our calculation hold, our finding likely indicates that the absorber has not reached an escape velocity, and is thus not a true outflow. Its kinematic properties (low line of sight velocity, high toroidal velocity component) are then more similar to Broad Line Region clouds in regular AGN \citep{Peterson+06}, although the ionization parameter of the absorber in \source\ is much higher.

Alternatively, perhaps this suggests that the black hole mass or emitting region size (in $R_{\rm G}$) is significantly smaller than we assumed, thus decreasing the necessary transverse velocity requirement and increasing the estimate of the maximum absorber distance from the X-ray source. Given this uncertain estimate of the emitting region size, we caution against using this result as a hard upper bound on the absorber location.

The ionized absorber could be launched directly from the newly-formed accretion disk of the TDE. It would probably have to originate in its outer part given the low projected outflow velocity of just a few 100s km/s. The launching mechanism is unclear but could be similar to the mechanism powering warm absorbers in regular AGN - possibly radiation line pressure \citep{Proga+00}, magnetic fields \citep{Fukumura+18} or thermal driving \citep{Waters+21}. Alternatively, the absorber could originate from shocked plasma in the stream-stream collisions of the TDE \citep{Jiang+16, Lu+20}. If this is the case, the photo-ionization calculation based on the assumption of photo-ionization equilibrium would not hold. However, the important limit on the location inferred from the time variability remains.

At this time, the origin of the ionized absorber remains unknown. Similar absorber detections in further TDEs are needed for a population study to resolve this issue. To our knowledge, \source\ is the first TDE to show a multi-phase low-velocity outflow in absorption. In particular, the low-ionization component has not been observed elsewhere. We note that \citet{Miller+15} found $3\sigma$ evidence for a second, redshifted ionized component in \ftnli, but that component is seen in emission and may form a P-Cygni profile with the primary ionized absorber.

Low-velocity ionized absorbers could be common among the TDE population, but no systematic, sample ionized outflow searches have been published to date. Such studies are challenging due to the data quality of the present high-spectral resolution TDE observations. Currently, only \xmm\ and \chandra\ gratings are capable of performing these studies. \source\ ($\sim8000$ source counts) and \ftnli\ ($>20000$ source counts) are two of the highest-quality datasets among the small number ($\lesssim10$) of TDEs with usable \xmm\ or \chandra\ grating spectra. Long exposure observations of bright sources, yielding at least a few thousand source counts in the gratings instruments, are necessary to perform a search for ionized outflow signatures.

Low-velocity outflow signatures in TDEs may also be recognized through spectral curvature in lower resolution CCD spectra, between 0.5 and 1.0 keV. However, it is challenging to confirm this interpretation with CCD-quality data alone. Similar spectral curvature can also be produced by more complex emission continua (double blackbody or blackbody+powerlaw versus a single blackbody), and other spectral features such as high-velocity absorbers (UFOs) in absorption.

New X-ray telescopes, with higher effective area and better spectral resolution in the soft X-ray band ($<1$ keV) are required. Two mission concepts are particularly well suited for observations of soft X-ray TDEs: the proposed X-ray probes Light Element Mapper \citep[LEM, ][]{Kraft+22} and Arcus \citep{Smith+20}. Either one would allow us to study the ionized absorption in TDEs in much greater detail (improved effective area and spectral resolution), and at greater distances (improved effective area), expanding the presently small population of TDEs with X-ray detected ionized absorbers.

\section{Conclusions} \label{sec:conclusions}

We analyze \xmm\ RGS and PN spectra of the tidal disruption event \source. The RGS spectrum reveals an array of narrow absorption lines, indicating the presence of an ionized absorber. Our conclusions are as follows:

\begin{itemize}

    \item The absorption lines cannot be described by a single photo-ionization phase, confirming a multi-phase nature of this plasma. There are at least 2 distinct phases: a highly ionized component with a column density of $\sim10^{23}$ \pcm\ and \logxi\ of 3.2, outflowing at 900 km/s, and a mildly ionized component with a column density of $\sim10^{21}$ \pcm, an ionization parameter of $\sim1.5$ and a velocity of 400 km/s.

    \item The ionized absorption varies in time during the single 60 ks \xmm\ exposure. The statistically strongest variation is observed in the ionization parameter of the mildly ionized component, but tentative variability is also detected in the highly ionized component.

    \item From the best-fitting parameters of the absorbers and their variability, we constrain the location of the ionized absorption to be as low as $\sim$400 $R_{\rm G}$ from the black hole. Consequently, we cannot be observing a (pc-scale) remnant outflow launched during previous black hole activity. The origin of the absorbers can be in a disk wind driven from the outer part of the TDE accretion disk, or in the shocked plasma created by stream-stream collisions of the tidally disrupted star.

\end{itemize}


\begin{acknowledgments}
 Support for this work was provided by the National Aeronautics and Space Administration through the Smithsonian Astrophysical Observatory (SAO) contract SV3-73016 to MIT for Support of the Chandra X-Ray Center and Science Instruments. PK and EK acknowledge support from NASA grants 80NSSC21K0872 and DD0-21125X. This work is based on observations obtained with \xmm, an ESA science mission funded by ESA Member States and USA (NASA).
\end{acknowledgments}

%

\vspace{5mm}
\facilities{\xmm
}


\software{SPEX \citep{Kaastra+96}, XSPEC \citep{Arnaud+96}, Veusz
          }






\bibliography{References}{}

\begin{thebibliography}{}
\expandafter\ifx\csname natexlab\endcsname\relax\def\natexlab#1{#1}\fi
\providecommand{\url}[1]{\href{#1}{#1}}
\providecommand{\dodoi}[1]{doi:~\href{http://doi.org/#1}{\nolinkurl{#1}}}
\providecommand{\doeprint}[1]{\href{http://ascl.net/#1}{\nolinkurl{http://ascl.net/#1}}}
\providecommand{\doarXiv}[1]{\href{https://arxiv.org/abs/#1}{\nolinkurl{https://arxiv.org/abs/#1}}}

\bibitem[{{Arnaud}(1996)}]{Arnaud+96}
{Arnaud}, K.~A. 1996, in Astronomical Society of the Pacific Conference Series,
  Vol. 101, Astronomical Data Analysis Software and Systems V, ed. G.~H.
  {Jacoby} \& J.~{Barnes}, 17

\bibitem[{{Blustin} {et~al.}(2005){Blustin}, {Page}, {Fuerst},
  {Branduardi-Raymont}, \& {Ashton}}]{Blustin+05}
{Blustin}, A.~J., {Page}, M.~J., {Fuerst}, S.~V., {Branduardi-Raymont}, G., \&
  {Ashton}, C.~E. 2005, \aap, 431, 111, \dodoi{10.1051/0004-6361:20041775}

\bibitem[{{Cash}(1979)}]{Cash+79}
{Cash}, W. 1979, \apj, 228, 939, \dodoi{10.1086/156922}

\bibitem[{{Chakraborty} {et~al.}(2021){Chakraborty}, {Kara}, {Masterson},
  {Giustini}, {Miniutti}, \& {Saxton}}]{Chakraborty+21}
{Chakraborty}, J., {Kara}, E., {Masterson}, M., {et~al.} 2021, \apjl, 921, L40,
  \dodoi{10.3847/2041-8213/ac313b}

\bibitem[{{Chartas} {et~al.}(2016){Chartas}, {Rhea}, {Kochanek}, {Dai},
  {Morgan}, {Blackburne}, {Chen}, {Mosquera}, \& {MacLeod}}]{Chartas+16}
{Chartas}, G., {Rhea}, C., {Kochanek}, C., {et~al.} 2016, Astronomische
  Nachrichten, 337, 356,
  \dodoi{10.1002/asna.20161231310.48550/arXiv.1509.05375}

\bibitem[{{Dai} {et~al.}(2018){Dai}, {McKinney}, {Roth}, {Ramirez-Ruiz}, \&
  {Miller}}]{Dai+18}
{Dai}, L., {McKinney}, J.~C., {Roth}, N., {Ramirez-Ruiz}, E., \& {Miller},
  M.~C. 2018, \apjl, 859, L20, \dodoi{10.3847/2041-8213/aab429}

\bibitem[{{Dai} {et~al.}(2010){Dai}, {Kochanek}, {Chartas}, {Koz{\l}owski},
  {Morgan}, {Garmire}, \& {Agol}}]{Dai+10}
{Dai}, X., {Kochanek}, C.~S., {Chartas}, G., {et~al.} 2010, \apj, 709, 278,
  \dodoi{10.1088/0004-637X/709/1/27810.48550/arXiv.0906.4342}

\bibitem[{{den Herder} {et~al.}(2001){den Herder}, {Brinkman}, {Kahn},
  {Branduardi-Raymont}, {Thomsen}, {Aarts}, {Audard}, {Bixler}, {den Boggende},
  {Cottam}, {Decker}, {Dubbeldam}, {Erd}, {Goulooze}, {G{\"u}del}, {Guttridge},
  {Hailey}, {Janabi}, {Kaastra}, {de Korte}, {van Leeuwen}, {Mauche},
  {McCalden}, {Mewe}, {Naber}, {Paerels}, {Peterson}, {Rasmussen}, {Rees},
  {Sakelliou}, {Sako}, {Spodek}, {Stern}, {Tamura}, {Tandy}, {de Vries},
  {Welch}, \& {Zehnder}}]{denHerder+01}
{den Herder}, J.~W., {Brinkman}, A.~C., {Kahn}, S.~M., {et~al.} 2001, \aap,
  365, L7, \dodoi{10.1051/0004-6361:20000058}

\bibitem[{{Detmers} {et~al.}(2010){Detmers}, {Kaastra}, {Costantini},
  {Verbunt}, {Cappi}, \& {de Vries}}]{Detmers+10}
{Detmers}, R.~G., {Kaastra}, J.~S., {Costantini}, E., {et~al.} 2010, \aap, 516,
  A61, \dodoi{10.1051/0004-6361/200913879}

\bibitem[{{Fukumura} {et~al.}(2018){Fukumura}, {Kazanas}, {Shrader}, {Behar},
  {Tombesi}, \& {Contopoulos}}]{Fukumura+18}
{Fukumura}, K., {Kazanas}, D., {Shrader}, C., {et~al.} 2018, \apj, 853, 40,
  \dodoi{10.3847/1538-4357/aaa3f610.48550/arXiv.1712.08181}

\bibitem[{{Gezari}(2021)}]{Gezari+21}
{Gezari}, S. 2021, \araa, 59, 21, \dodoi{10.1146/annurev-astro-111720-030029}

\bibitem[{{Hammerstein} {et~al.}(2023){Hammerstein}, {van Velzen}, {Gezari},
  {Cenko}, {Yao}, {Ward}, {Frederick}, {Villanueva}, {Somalwar}, {Graham},
  {Kulkarni}, {Stern}, {Andreoni}, {Bellm}, {Dekany}, {Dhawan}, {Drake},
  {Fremling}, {Gatkine}, {Groom}, {Ho}, {Kasliwal}, {Karambelkar}, {Kool},
  {Masci}, {Medford}, {Perley}, {Purdum}, {van Roestel}, {Sharma}, {Sollerman},
  {Taggart}, \& {Yan}}]{Hammerstein+23}
{Hammerstein}, E., {van Velzen}, S., {Gezari}, S., {et~al.} 2023, \apj, 942, 9,
  \dodoi{10.3847/1538-4357/aca283}

\bibitem[{{HI4PI Collaboration} {et~al.}(2016){HI4PI Collaboration}, {Ben
  Bekhti}, {Fl{\"o}er}, {Keller}, {Kerp}, {Lenz}, {Winkel}, {Bailin},
  {Calabretta}, {Dedes}, {Ford}, {Gibson}, {Haud}, {Janowiecki}, {Kalberla},
  {Lockman}, {McClure-Griffiths}, {Murphy}, {Nakanishi}, {Pisano}, \&
  {Staveley-Smith}}]{HI4PI+16}
{HI4PI Collaboration}, {Ben Bekhti}, N., {Fl{\"o}er}, L., {et~al.} 2016, \aap,
  594, A116

\bibitem[{{Hinkle}(2022)}]{Hinkle+22}
{Hinkle}, J.~T. 2022, arXiv e-prints, arXiv:2210.15681,
  \dodoi{10.48550/arXiv.2210.15681}

\bibitem[{{Jansen} {et~al.}(2001){Jansen}, {Lumb}, {Altieri}, {Clavel}, {Ehle},
  {Erd}, {Gabriel}, {Guainazzi}, {Gondoin}, {Much}, {Munoz}, {Santos},
  {Schartel}, {Texier}, \& {Vacanti}}]{Jansen+01}
{Jansen}, F., {Lumb}, D., {Altieri}, B., {et~al.} 2001, \aap, 365, L1,
  \dodoi{10.1051/0004-6361:20000036}

\bibitem[{{Jiang} {et~al.}(2016){Jiang}, {Guillochon}, \& {Loeb}}]{Jiang+16}
{Jiang}, Y.-F., {Guillochon}, J., \& {Loeb}, A. 2016, \apj, 830, 125,
  \dodoi{10.3847/0004-637X/830/2/125}

\bibitem[{{Kaastra} {et~al.}(1996){Kaastra}, {Mewe}, \&
  {Nieuwenhuijzen}}]{Kaastra+96}
{Kaastra}, J.~S., {Mewe}, R., \& {Nieuwenhuijzen}, H. 1996, in UV and X-ray
  Spectroscopy of Astrophysical and Laboratory Plasmas, 411--414

\bibitem[{{Kara} {et~al.}(2018){Kara}, {Dai}, {Reynolds}, \&
  {Kallman}}]{Kara+18}
{Kara}, E., {Dai}, L., {Reynolds}, C.~S., \& {Kallman}, T. 2018, \mnras, 474,
  3593, \dodoi{10.1093/mnras/stx3004}

\bibitem[{{Kara} {et~al.}(2016){Kara}, {Miller}, {Reynolds}, \&
  {Dai}}]{Kara+16}
{Kara}, E., {Miller}, J.~M., {Reynolds}, C., \& {Dai}, L. 2016, \nat, 535, 388,
  \dodoi{10.1038/nature18007}

\bibitem[{{Kosec} {et~al.}(2020){Kosec}, {Fabian}, {Pinto}, {Walton}, {Dyda},
  \& {Reynolds}}]{Kosec+20a}
{Kosec}, P., {Fabian}, A.~C., {Pinto}, C., {et~al.} 2020, \mnras, 491, 3730,
  \dodoi{10.1093/mnras/stz3200}

\bibitem[{{Kosec} {et~al.}(2018){Kosec}, {Pinto}, {Walton}, {Fabian},
  {Bachetti}, {Brightman}, {F{\"u}rst}, \& {Grefenstette}}]{Kosec+18b}
{Kosec}, P., {Pinto}, C., {Walton}, D.~J., {et~al.} 2018, \mnras, 479, 3978,
  \dodoi{10.1093/mnras/sty1626}

\bibitem[{{Kraft} {et~al.}(2022){Kraft}, {Markevitch}, {Kilbourne}, {Adams},
  {Akamatsu}, {Ayromlou}, {Bandler}, {Bennett}, {Bhardwaj}, {Biffi},
  {Bodewits}, {Bogdan}, {Bonamente}, {Borgani}, {Branduardi-Raymont},
  {Bregman}, {Burchett}, {Cann}, {Carter}, {Chakraborty}, {Churazov}, {Crain},
  {Cumbee}, {Dave}, {DiPirro}, {Dolag}, {Bertrand Doriese}, {Drake}, {Dunn},
  {Eckart}, {Eckert}, {Ettori}, {Forman}, {Galeazzi}, {Gall}, {Gatuzz}, {Hell},
  {Hodges-Kluck}, {Jackman}, {Jahromi}, {Jennings}, {Jones}, {Kaaret},
  {Kavanagh}, {Kelley}, {Khabibullin}, {Kim}, {Koutroumpa}, {Kovacs}, {Kuntz},
  {Lin}, {Lau}, {Lee}, {Leutenegger}, {Lisse}, {Lovisari}, {McCammon},
  {McEntee}, {Mernier}, {Miller}, {Nagai}, {Negro}, {Nelson}, {Ness}, {Nulsen},
  {Ogorzalek}, {Oppenheimer}, {Oskinova}, {Patnaude}, {Pfeifle}, {Pillepich},
  {Plucinsky}, {Pooley}, {Porter}, {Randall}, {Rasia}, {Raymond}, {Ruszkowski},
  {Sakai}, {Sarkar}, {Sasaki}, {Sato}, {Schellenberger}, {Schaye},
  {Simionescu}, {Smith}, {Steiner}, {Stern}, {Su}, {Sun}, {Tremblay}, {Truong},
  {Tutt}, {Veilleux}, {Vikhlinin}, {Vladutescu-Zopp}, {Vogelsberger}, {Walker},
  {Weaver}, {Weigt}, {Werk}, {Werner}, {Wolk}, {Zhang}, {Zhang}, {Zhuravleva},
  \& {ZuHone}}]{Kraft+22}
{Kraft}, R., {Markevitch}, M., {Kilbourne}, C., {et~al.} 2022, arXiv e-prints,
  arXiv:2211.09827, \dodoi{10.48550/arXiv.2211.09827}

\bibitem[{{Laha} {et~al.}(2014){Laha}, {Guainazzi}, {Dewangan}, {Chakravorty},
  \& {Kembhavi}}]{Laha+14}
{Laha}, S., {Guainazzi}, M., {Dewangan}, G.~C., {Chakravorty}, S., \&
  {Kembhavi}, A.~K. 2014, \mnras, 441, 2613, \dodoi{10.1093/mnras/stu669}

\bibitem[{{Lu} \& {Bonnerot}(2020)}]{Lu+20}
{Lu}, W., \& {Bonnerot}, C. 2020, \mnras, 492, 686,
  \dodoi{10.1093/mnras/stz3405}

\bibitem[{{Masterson} {et~al.}(2022){Masterson}, {Kara}, {Ricci},
  {Garc{\'\i}a}, {Fabian}, {Pinto}, {Kosec}, {Remillard}, {Loewenstein},
  {Trakhtenbrot}, \& {Arcavi}}]{Masterson+22}
{Masterson}, M., {Kara}, E., {Ricci}, C., {et~al.} 2022, \apj, 934, 35,
  \dodoi{10.3847/1538-4357/ac76c0}

\bibitem[{{Mehdipour} {et~al.}(2016){Mehdipour}, {Kaastra}, \&
  {Kallman}}]{Mehdipour+16}
{Mehdipour}, M., {Kaastra}, J.~S., \& {Kallman}, T. 2016, \aap, 596, A65

\bibitem[{{Miller} {et~al.}(2015){Miller}, {Kaastra}, {Miller}, {Reynolds},
  {Brown}, {Cenko}, {Drake}, {Gezari}, {Guillochon}, {Gultekin}, {Irwin},
  {Levan}, {Maitra}, {Maksym}, {Mushotzky}, {O'Brien}, {Paerels}, {de Plaa},
  {Ramirez-Ruiz}, {Strohmayer}, \& {Tanvir}}]{Miller+15}
{Miller}, J.~M., {Kaastra}, J.~S., {Miller}, M.~C., {et~al.} 2015, \nat, 526,
  542

\bibitem[{{Miniutti} {et~al.}(2019){Miniutti}, {Saxton}, {Giustini},
  {Alexander}, {Fender}, {Heywood}, {Monageng}, {Coriat}, {Tzioumis}, {Read},
  {Knigge}, {Gandhi}, {Pretorius}, \& {Ag{\'\i}s-Gonz{\'a}lez}}]{Miniutti+19}
{Miniutti}, G., {Saxton}, R.~D., {Giustini}, M., {et~al.} 2019, \nat, 573, 381,
  \dodoi{10.1038/s41586-019-1556-x}

\bibitem[{{Morgan} {et~al.}(2008){Morgan}, {Kochanek}, {Dai}, {Morgan}, \&
  {Falco}}]{Morgan+08}
{Morgan}, C.~W., {Kochanek}, C.~S., {Dai}, X., {Morgan}, N.~D., \& {Falco},
  E.~E. 2008, \apj, 689, 755, \dodoi{10.1086/59276710.48550/arXiv.0802.1210}

\bibitem[{{Nicholl} {et~al.}(2020){Nicholl}, {Wevers}, {Oates}, {Alexander},
  {Leloudas}, {Onori}, {Jerkstrand}, {Gomez}, {Campana}, {Arcavi},
  {Charalampopoulos}, {Gromadzki}, {Ihanec}, {Jonker}, {Lawrence}, {Mandel},
  {Schulze}, {Short}, {Burke}, {McCully}, {Hiramatsu}, {Howell}, {Pellegrino},
  {Abbot}, {Anderson}, {Berger}, {Blanchard}, {Cannizzaro}, {Chen},
  {Dennefeld}, {Galbany}, {Gonz{\'a}lez-Gait{\'a}n}, {Hosseinzadeh}, {Inserra},
  {Irani}, {Kuin}, {M{\"u}ller-Bravo}, {Pineda}, {Ross}, {Roy}, {Smartt},
  {Smith}, {Tucker}, {Wyrzykowski}, \& {Young}}]{Nicholl+20}
{Nicholl}, M., {Wevers}, T., {Oates}, S.~R., {et~al.} 2020, \mnras, 499, 482,
  \dodoi{10.1093/mnras/staa2824}

\bibitem[{{Ohsuga} \& {Mineshige}(2011)}]{Ohsuga+11}
{Ohsuga}, K., \& {Mineshige}, S. 2011, \apj, 736, 2,
  \dodoi{10.1088/0004-637X/736/1/2}

\bibitem[{{Ohsuga} {et~al.}(2009){Ohsuga}, {Mineshige}, {Mori}, \&
  {Kato}}]{Ohsuga+09}
{Ohsuga}, K., {Mineshige}, S., {Mori}, M., \& {Kato}, Y. 2009, \pasj, 61, L7,
  \dodoi{10.1093/pasj/61.3.L7}

\bibitem[{{Ohsuga} {et~al.}(2005){Ohsuga}, {Mori}, {Nakamoto}, \&
  {Mineshige}}]{Ohsuga+05}
{Ohsuga}, K., {Mori}, M., {Nakamoto}, T., \& {Mineshige}, S. 2005, \apj, 628,
  368, \dodoi{10.1086/430728}

\bibitem[{{Partington} {et~al.}(2023){Partington}, {Cackett}, {Kara}, {Kriss},
  {Barth}, {De Rosa}, {Homayouni}, {Horne}, {Landt}, {Zoghbi}, {Edelson},
  {Arav}, {Boizelle}, {Bentz}, {Brotherton}, {Byun}, {Dalla Bont{\`a}},
  {Dehghanian}, {Du}, {Fian}, {Filippenko}, {Gelbord}, {Goad}, {Gonz{\'a}lez
  Buitrago}, {Grier}, {Hall}, {Hu}, {Ili{\'c}}, {Joner}, {Kaspi}, {Kochanek},
  {Korista}, {Kova{\v{c}}evi{\'c}}, {Kynoch}, {McLane}, {Mehdipour}, {Miller},
  {Panagiotou}, {Plesha}, {Popovi{\'c}}, {Proga}, {Rogantini},
  {Storchi-Bergmann}, {Sanmartim}, {Siebert}, {Vestergaard}, {Ward}, {Waters},
  \& {Zaidouni}}]{Partington+23}
{Partington}, E.~R., {Cackett}, E.~M., {Kara}, E., {et~al.} 2023, \apj, 947, 2,
  \dodoi{10.3847/1538-4357/acbf44}

\bibitem[{{Peterson}(2006)}]{Peterson+06}
{Peterson}, B.~M. 2006, in Physics of Active Galactic Nuclei at all Scales, ed.
  D.~{Alloin}, Vol. 693, 77, \dodoi{10.1007/3-540-34621-X_3}

\bibitem[{{Pinto} {et~al.}(2016){Pinto}, {Middleton}, \& {Fabian}}]{Pinto+16}
{Pinto}, C., {Middleton}, M.~J., \& {Fabian}, A.~C. 2016, \nat, 533, 64

\bibitem[{{Pinto} {et~al.}(2021){Pinto}, {Soria}, {Walton}, {D'A{\`\i}},
  {Pintore}, {Kosec}, {Alston}, {Fuerst}, {Middleton}, {Roberts}, {Del Santo},
  {Barret}, {Ambrosi}, {Robba}, {Earnshaw}, \& {Fabian}}]{Pinto+21}
{Pinto}, C., {Soria}, R., {Walton}, D.~J., {et~al.} 2021, \mnras, 505, 5058,
  \dodoi{10.1093/mnras/stab1648}

\bibitem[{{Pounds} {et~al.}(2003){Pounds}, {Reeves}, {King}, {Page}, {O'Brien},
  \& {Turner}}]{Pounds+03}
{Pounds}, K.~A., {Reeves}, J.~N., {King}, A.~R., {et~al.} 2003, \mnras, 345,
  705, \dodoi{10.1046/j.1365-8711.2003.07006.x}

\bibitem[{{Proga} {et~al.}(2000){Proga}, {Stone}, \& {Kallman}}]{Proga+00}
{Proga}, D., {Stone}, J.~M., \& {Kallman}, T.~R. 2000, \apj, 543, 686,
  \dodoi{10.1086/31715410.48550/arXiv.astro-ph/0005315}

\bibitem[{{Rees}(1988)}]{Rees+88}
{Rees}, M.~J. 1988, \nat, 333, 523, \dodoi{10.1038/333523a0}

\bibitem[{{Ricci} {et~al.}(2020){Ricci}, {Kara}, {Loewenstein}, {Trakhtenbrot},
  {Arcavi}, {Remillard}, {Fabian}, {Gendreau}, {Arzoumanian}, {Li}, {Ho},
  {MacLeod}, {Cackett}, {Altamirano}, {Gandhi}, {Kosec}, {Pasham}, {Steiner},
  \& {Chan}}]{Ricci+20}
{Ricci}, C., {Kara}, E., {Loewenstein}, M., {et~al.} 2020, \apjl, 898, L1,
  \dodoi{10.3847/2041-8213/ab91a1}

\bibitem[{{Sanfrutos} {et~al.}(2013){Sanfrutos}, {Miniutti},
  {Ag{\'\i}s-Gonz{\'a}lez}, {Fabian}, {Miller}, {Panessa}, \&
  {Zoghbi}}]{Sanfrutos+13}
{Sanfrutos}, M., {Miniutti}, G., {Ag{\'\i}s-Gonz{\'a}lez}, B., {et~al.} 2013,
  \mnras, 436, 1588, \dodoi{10.1093/mnras/stt167510.48550/arXiv.1309.1092}

\bibitem[{{Saxton} {et~al.}(2020){Saxton}, {Komossa}, {Auchettl}, \&
  {Jonker}}]{Saxton+20}
{Saxton}, R., {Komossa}, S., {Auchettl}, K., \& {Jonker}, P.~G. 2020, \ssr,
  216, 85, \dodoi{10.1007/s11214-020-00708-4}

\bibitem[{{Sazonov} {et~al.}(2021){Sazonov}, {Gilfanov}, {Medvedev}, {Yao},
  {Khorunzhev}, {Semena}, {Sunyaev}, {Burenin}, {Lyapin}, {Meshcheryakov},
  {Uskov}, {Zaznobin}, {Postnov}, {Dodin}, {Belinski}, {Cherepashchuk},
  {Eselevich}, {Dodonov}, {Grokhovskaya}, {Kotov}, {Bikmaev}, {Zhuchkov},
  {Gumerov}, {van Velzen}, \& {Kulkarni}}]{Sazonov+21}
{Sazonov}, S., {Gilfanov}, M., {Medvedev}, P., {et~al.} 2021, \mnras, 508,
  3820, \dodoi{10.1093/mnras/stab2843}

\bibitem[{{Shakura} \& {Sunyaev}(1973)}]{Shakura+73}
{Shakura}, N.~I., \& {Sunyaev}, R.~A. 1973, \aap, 24, 337

\bibitem[{{Smith}(2020)}]{Smith+20}
{Smith}, R.~K. 2020, in Society of Photo-Optical Instrumentation Engineers
  (SPIE) Conference Series, Vol. 11444, Society of Photo-Optical
  Instrumentation Engineers (SPIE) Conference Series, 114442C,
  \dodoi{10.1117/12.2576047}

\bibitem[{{Stanek}(2020)}]{Stanek+20}
{Stanek}, K.~Z. 2020, Transient Name Server Discovery Report, 2020-3850, 1

\bibitem[{{Str{\"u}der} {et~al.}(2001){Str{\"u}der}, {Briel}, {Dennerl},
  {Hartmann}, {Kendziorra}, {Meidinger}, {Pfeffermann}, {Reppin}, {Aschenbach},
  {Bornemann}, {Br{\"a}uninger}, {Burkert}, {Elender}, {Freyberg}, {Haberl},
  {Hartner}, {Heuschmann}, {Hippmann}, {Kastelic}, {Kemmer}, {Kettenring},
  {Kink}, {Krause}, {M{\"u}ller}, {Oppitz}, {Pietsch}, {Popp}, {Predehl},
  {Read}, {Stephan}, {St{\"o}tter}, {Tr{\"u}mper}, {Holl}, {Kemmer}, {Soltau},
  {St{\"o}tter}, {Weber}, {Weichert}, {von Zanthier}, {Carathanassis}, {Lutz},
  {Richter}, {Solc}, {B{\"o}ttcher}, {Kuster}, {Staubert}, {Abbey}, {Holland},
  {Turner}, {Balasini}, {Bignami}, {La Palombara}, {Villa}, {Buttler},
  {Gianini}, {Lain{\'e}}, {Lumb}, \& {Dhez}}]{Struder+01}
{Str{\"u}der}, L., {Briel}, U., {Dennerl}, K., {et~al.} 2001, \aap, 365, L18,
  \dodoi{10.1051/0004-6361:20000066}

\bibitem[{{Takeuchi} {et~al.}(2013){Takeuchi}, {Ohsuga}, \&
  {Mineshige}}]{Takeuchi+13}
{Takeuchi}, S., {Ohsuga}, K., \& {Mineshige}, S. 2013, \pasj, 65, 88,
  \dodoi{10.1093/pasj/65.4.88}

\bibitem[{{Tombesi} {et~al.}(2011){Tombesi}, {Cappi}, {Reeves}, {Palumbo},
  {Braito}, \& {Dadina}}]{Tombesi+11}
{Tombesi}, F., {Cappi}, M., {Reeves}, J.~N., {et~al.} 2011, \apj, 742, 44,
  \dodoi{10.1088/0004-637X/742/1/44}

\bibitem[{{Tombesi} {et~al.}(2010){Tombesi}, {Cappi}, {Reeves}, {Palumbo},
  {Yaqoob}, {Braito}, \& {Dadina}}]{Tombesi+10}
---. 2010, \aap, 521, A57

\bibitem[{{van Velzen} {et~al.}(2021){van Velzen}, {Gezari}, {Hammerstein},
  {Roth}, {Frederick}, {Ward}, {Hung}, {Cenko}, {Stein}, {Perley}, {Taggart},
  {Foley}, {Sollerman}, {Blagorodnova}, {Andreoni}, {Bellm}, {Brinnel}, {De},
  {Dekany}, {Feeney}, {Fremling}, {Giomi}, {Golkhou}, {Graham}, {Ho},
  {Kasliwal}, {Kilpatrick}, {Kulkarni}, {Kupfer}, {Laher}, {Mahabal}, {Masci},
  {Miller}, {Nordin}, {Riddle}, {Rusholme}, {van Santen}, {Sharma}, {Shupe}, \&
  {Soumagnac}}]{vanVelzen+21}
{van Velzen}, S., {Gezari}, S., {Hammerstein}, E., {et~al.} 2021, \apj, 908, 4,
  \dodoi{10.3847/1538-4357/abc258}

\bibitem[{{Waters} {et~al.}(2021){Waters}, {Proga}, \& {Dannen}}]{Waters+21}
{Waters}, T., {Proga}, D., \& {Dannen}, R. 2021, \apj, 914, 62,
  \dodoi{10.3847/1538-4357/abfbe610.48550/arXiv.2101.09273}

\end{thebibliography}
\bibliographystyle{aasjournal}



\end{document}